\documentclass[%
reprint,
superscriptaddress,
amsmath,amssymb,
aps,
prb,
]{revtex4-2}
\usepackage{graphicx}
\usepackage{dcolumn}
\usepackage{bm}
\usepackage[mathlines]{lineno}
\usepackage[colorlinks,
linkcolor=blue,
anchorcolor=blue, 
citecolor=blue,
urlcolor=blue,
]{hyperref}
\usepackage{hyperref} 
\usepackage{siunitx}
\DeclareSIUnit\gauss{G}
\usepackage{multirow} 
\usepackage{sidecap}
\usepackage[version=4]{mhchem}

\usepackage{xspace}

\newcommand{\tplus}{\ensuremath{^{3+}}\xspace}
\newcommand{\erg}{\ensuremath{^4}I\ensuremath{_{15/2}}\xspace}
\newcommand{\ere}{\ensuremath{^4}I\ensuremath{_{13/2}}\xspace}
\newcommand{\yso}{Y\ensuremath{_2}SiO\ensuremath{_5}\xspace}
\begin{document}
\title{The Zeeman and hyperfine interactions of a single $^{167}$Er\tplus ion in Si}

\affiliation{CAS Key Laboratory of Microscale Magnetic Resonance and School of Physical Sciences, University of Science and Technology of China, Hefei 230026, China}
\affiliation{Hefei National Laboratory, Hefei 230088, China}
\affiliation{CAS Center for Excellence in Quantum Information and Quantum Physics, University of Science and Technology of China, Hefei 230026, China}

\author{Jiliang Yang}
\author{Wenda Fan}
\author{Yangbo Zhang}
\author{Changkui Duan}
\affiliation{CAS Key Laboratory of Microscale Magnetic Resonance and School of Physical Sciences, University of Science and Technology of China, Hefei 230026, China}
\affiliation{CAS Center for Excellence in Quantum Information and Quantum Physics, University of Science and Technology of China, Hefei 230026, China}
\author{Gabriele G. de Boo}
\affiliation{Centre of Excellence for Quantum Computation and Communication Technology, School of Physics, University of New South Wales, NSW 2052, Australia}
\author{Rose L. Ahlefeldt}
\affiliation{Centre of Excellence for Quantum Computation and Communication Technology, Research School of Physics, Australian National University, ACT 0200, Australia}
\author{Jevon J. Longdell}
\affiliation{The Dodd-Walls Centre for Photonic and Quantum Technologies, Department of Physics, University of Otago, Dunedin 9016, New Zealand}
\author{Brett C. Johnson}
\affiliation{Centre of Excellence for Quantum Computation and Communication Technology, School of Engineering, RMIT University, Victoria 3001, Australia}
\affiliation{Centre of Excellence for Quantum Computation and Communication Technology, School of Physics, University of Melbourne, Victoria 3010, Australia}
\author{Jeffrey C. McCallum}
\affiliation{Centre of Excellence for Quantum Computation and Communication Technology, School of Physics, University of Melbourne, Victoria 3010, Australia}
\author{Matthew J. Sellars}
\affiliation{Centre of Excellence for Quantum Computation and Communication Technology, Research School of Physics, Australian National University, ACT 0200, Australia}
\author{Sven Rogge}
\affiliation{Centre of Excellence for Quantum Computation and Communication Technology, School of Physics, University of New South Wales, NSW 2052, Australia}
\author{Chunming Yin}
\email{Chunming@ustc.edu.cn}
\author{Jiangfeng Du}
\affiliation{CAS Key Laboratory of Microscale Magnetic Resonance and School of Physical Sciences, University of Science and Technology of China, Hefei 230026, China}
\affiliation{Hefei National Laboratory, Hefei 230088, China}
\affiliation{CAS Center for Excellence in Quantum Information and Quantum Physics, University of Science and Technology of China, Hefei 230026, China}

\date{\today}
\begin{abstract}
Er-doped Si is a promising candidate for quantum information applications due to its telecom wavelength optical transition and its compatibility with Si nanofabrication technologies. Recent spectroscopic studies based on photoluminescence excitation have shown multiple well-defined lattice sites that Er occupies in Si. Here we report the first measurement of the Zeeman and hyperfine tensors of a single $^{167}$Er\tplus ion in Si. All the obtained tensors are highly anisotropic with the largest value principal axes aligning in nearly the same direction, and the trace of the lowest crystal field level $\mathbf{g}$-tensor is \num{17.78+-0.40}. The results indicate that this specific Er site is likely to be a distorted cubic site that exhibits monoclinic (C$_1$) symmetry. Finally, zero first-order-Zeeman (ZEFOZ) fields are identified for this site and could be used to reduce decoherence of hyperfine spin states in future experiments.
\end{abstract}

\maketitle
\section{introduction}
Er\tplus ions have been widely used in classical optical devices and networks thanks to the \erg-\ere optical transition at telecom wavelength. Recent studies on Er\tplus ions have demonstrated new prospects for quantum optics and quantum computing with direct optical access to telecom fibre networks and Si photonic cavities. Er\tplus is particularly attractive for quantum applications because of the long coherence time available on hyperfine states. For example, hyperfine coherence time can exceed \SI{1}{\s} for Er\tplus ions in \yso\cite{Coherence_18}. Further enhancement can be achieved by utilising hyperfine transitions at a zero first-order-Zeeman (ZEFOZ) field that are protected from magnetic field fluctuations\cite{zhong_optically_2015}. In addition, strong cavity coupling and coherent control of a single Er\tplus ion have been demonstrated by combining a Si cavity with an Er-doped \yso crystal\cite{dibos_atomic_2018,chen_parallel_2020}.

Si would be an obvious choice of host materials for Er\tplus ions due to the well-established Si nanofabrication technologies. Also, the nuclear-spin free environment provided by isotopically purified $^{28}$Si can significantly suppress decoherence due to nuclear spins in the host crystal. One challenge is that Er\tplus ions can form a range of sites in Si in contrast to only two crystallographic sites that Er\tplus ions occupy in \yso\cite{Chen_hyperfine_18}. By choosing a suitable annealing condition and Er concentration, a cubic Er site can become the dominant site in float-zone (FZ) Si with low impurity levels\cite{przybylinska_optically_1996}. The typical structure of its photoluminescence (PL) spectra includes an optical transition at \SI{195.01}{\THz} between its lowest levels of the \erg and \ere states. While complex spectra appear in Er-doped Si under different annealing conditions or in the presence of other dopants or implantation induced defects\cite{przybylinska_local_1995}, the spectral structure of this cubic site is still observable. Analysis of the fine structure suggests the Er\tplus ion sees a tetrahedral crystal field\cite{tang_characteristics_1989,przybylinska_optically_1996}.  Also, existence of a tetrahedral interstitial Er site in Si was confirmed by emission channelling studies\cite{Wahl_TdEr_97}, although there was no direct experimental link between this result and the earlier measurements. These findings are in agreement with several theoretical studies that the tetrahedral interstitial site is the most stable site for Er in Si\cite{needels_erbium_1993,wan_role_1998,hashimoto_determination_2001,prezzi_electrical_2005}, but other studies found a tetrahedral substitutional site\cite{raffa_equilibrium_2002,delerue_description_1991} to be more stable or a hexagonal interstitial site when oxygen is involved\cite{raffa_equilibrium_2002,wan_role_1998}.

The most direct evidence of a cubic site would be an isotropic Zeeman $\mathbf{g}$-tensor, but the $\mathbf{g}$-tensor of the cubic Er site in Si has not been determined due to the complex spectra and limited resolution from PL or electron paramagnetic resonance (EPR) measurements. Another characteristic property is the trace of the Z$_1$ $\mathbf{g}$-tensor, where Z$_1$ represents the lowest crystal field level of the \erg ground state. Crystal field levels of a cubic site can be either $\Gamma_8$ quartet states or doublet states of $\Gamma_6$ or $\Gamma_7$, but only $\Gamma_6$ and $\Gamma_7$ Z$_1$ levels have been experimentally identified in Er-doped crystals\cite{Maat_Zeeman_01}. The trace of the Z$_1$ $\mathbf{g}$-tensor for $\Gamma_6$ and $\Gamma_7$ is calculated to be \num{20.4} and \num{18.0} based on crystal field theory, and the measured values in two cubic crystals are on average \num{20.27+-0.12} and \num{17.79+-0.03}, respectively\cite{Maat_Zeeman_01}. The small deviation from the calculated value is usually explained by interactions with higher-lying energy levels or the effects of covalency\cite{watts_paramagnetic-resonance_1968,carey_electron_1999}. Er\tplus ions can also occupy a slightly distorted cubic site where the crystal field distortion is small compared to the cubic crystal field. As a result, the distortion on the trace of the Z$_1$ $\mathbf{g}$-tensor is expected to be smaller than \num{0.3}\cite{carey_electron_1999} and the cubic crystal field approximation is still valid\cite{carey_state_2009}.

\begin{table}
	\caption{\label{tab:table1}List of devices in which Er\tplus transitions were observed by detecting the ionisation of a single trap.}
	\begin{tabular}{l l l}
		 \hline
		 \hline
          Device & Transitions (THz) & Note\\ 
          \hline
          1 &194.05, 195.03, 195.07, 196.02 &\\
          2 &195.03 &\\
		  3 &195.04, 195.94 &\\
		  4 &195.04, 195.99 & The present work\\
		  5 &195.07 &\\
		  6 &195.12 &\\
		  7 &195.15 &\\ 
		  8 &195.35 & Ref.\cite{hu_er_2022}\\
		 \hline
		 \hline
    \end{tabular}
\end{table}

Recent high-resolution photoluminescence excitation (PLE) measurements on Si waveguides\cite{Weiss_21, gritsch2021narrow} and bulk Si\cite{Berkman_Er_21} have both shown multiple well-defined Er sites. The inhomogeneous linewidths are of the order of \SI{1}{\GHz}, which is comparable to that of Er:\yso\tplus, and some sites show homogeneous linewidths below \SI{1}{\MHz}. These studies reinforce the promise of Er in Si, but for applications where long coherence times are required, extensive studies of the Zeeman and hyperfine interactions of the Er sites in Si are also required. The Zeeman interaction defines the preferred magnetic field direction to suppress electron spin relaxation due to spin-lattice coupling. Further, hyperfine transitions at a ZEFOZ field are protected from magnetic field fluctuations and thus allow extended spin coherence times\cite{reducing_12}. The Zeeman splitting can be observed for some Er sites using PLE, but the resolution is limited by the $\sim$\SI{1}{\GHz} inhomogeneous linewidth and may be further limited by multiple site orientations\cite{Berkman_Er_21}. 

\begin{figure*}[t]
    \includegraphics[width=\textwidth]{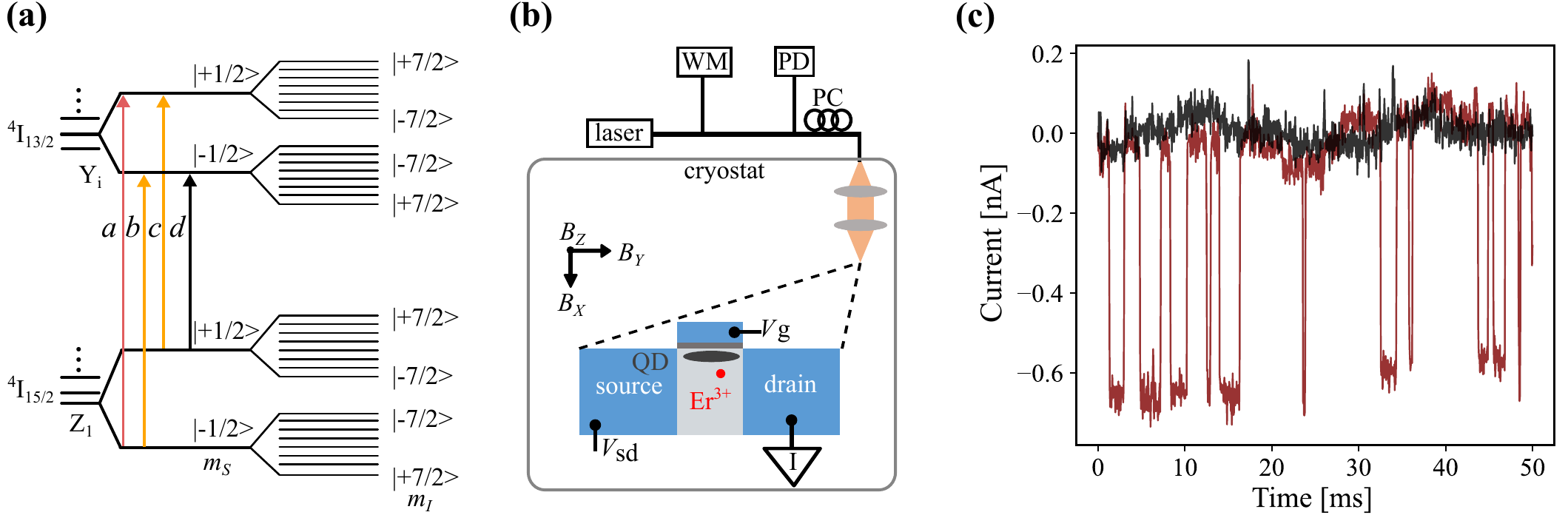}
    \caption{Spectroscopic measurement of a $^{167}$Er\tplus ion. (a) A generic energy level diagram of a $^{167}$Er\tplus ion in a high magnetic field. $m_S$ and $m_I$ represent the spin projection quantum numbers of the electronic and hyperfine spin states, respectively. (b) A schematic of the FinFET device and its electrical and optical connections. The laser beam is focused on the channel region of the device doped with $^{167}$Er\tplus ions. (c) Two typical current-time traces measured under resonant (red) and non-resonant (black) illumination.}
    \label{fig:fig1}
\end{figure*}

The inhomogeneous broadening can be avoided by looking at only one ion. The hybrid electrical/optical single ion detection enables the high spectral resolution study of the Zeeman and hyperfine interactions on a single ion level\cite{Yin_13,de_boo_high-resolution_2020}. Recently, a time-resolved single ion detection technique was demonstrated\cite{hu_er_2022} by detecting the ionisation of a single trap, as opposed to multiple traps in the previously used time-averaged measurements\cite{Yin_13,zhang_single_2019}. The time-resolved detection has the advantage of allowing a wider range of experimental conditions and faster repetition of the single ion detection than the time-averaged detection\cite{hu_er_2022}. While the earlier studies with time-averaged detection revealed a large number of optical transitions from \SIrange{193.5}{197.0}{\THz}\cite{Berkman_Er_21}, optical transitions at approximately \SI{195.04}{\THz} are frequently observed in devices showing single trap ionisation, as shown in Table~\ref{tab:table1}.

Here we investigate the Zeeman and hyperfine interactions of a single $^{167}$Er\tplus ion in Si. This Er site has a zero field optical transition frequency of \SI{195036.7}{\GHz} (\SI{1537.107}{\nm}) which closely matches the previously reported cubic site in Si\cite{przybylinska_optically_1996}. A spin Hamiltonian model is used to fit the spectra from field rotation measurements. The fitting gives the Zeeman interaction $\mathbf{g}$-tensors and the hyperfine interaction $\mathbf{A}$-tensors for the lowest crystal field level in the \erg manifold and a crystal field level in the \ere manifold of the $^{167}$Er\tplus ion.

\section{Spin Hamiltonian model}
Er forms a trivalent state in most semiconductors with a 4f$^{11}$ electronic configuration. The telecom wavelength optical transitions arise from the \erg ground state multiplet and the \ere first optically excited state. 
Under the action of a crystal field, the $2J+1$ degeneracy in these levels is partially or wholly lifted. For example, the \erg state splits into five levels for the special case of the cubic site and eight levels for low symmetry sites.

Figure~\ref{fig:fig1}(a) shows a generic energy level diagram of the low symmetry Er site investigated in this work. The crystal field levels of \erg (\ere) are labelled as Z$_1$ up to Z$_8$ (Y$_1$ up to Y$_7$) in order from the lowest energy to the highest. Each crystal field level has a two-fold electron spin degeneracy (Kramers doublet) which can be lifted by a magnetic field. 
For a $^{167}$Er\tplus ion with a nuclear spin of $I = 7/2$, each crystal field doublet splits into 16 hyperfine sublevels. The Zeeman and hyperfine splittings of a crystal field level can typically be described in low field by a spin Hamiltonian, \cite{Guillot_hyperfine_06,crystalmodel_19,Chen_hyperfine_18} 
\begin{equation}
	{H=\mu_{e} \mathbf{B}\cdot\mathbf{g}\cdot\mathbf{S}+\mathbf{I}\cdot\mathbf{A}\cdot\mathbf{S}+\mathbf{I}\cdot\mathbf{Q}\cdot\mathbf{I}-\mu_{n}g_{n}\mathbf{B}\cdot\mathbf{I}}
	\label{eq:eq1}
\end{equation}
where $\mathbf{B}$ is the external magnetic field, $\mathbf{S}$ is an effective spin vector with $S = 1/2$, $\mathbf{I}$ is a nuclear spin vector with a value of $I = 7/2$, $\mathbf{g}$, $\mathbf{A}$ and $\mathbf{Q}$ are the Zeeman, hyperfine and quadrupole tensors, $\mu_{e}$ and $\mu_{n}$ are the Bohr and nuclear magneton, respectively, and $g_{n}\sim-$0.1618 is the nuclear $g$ factor.
\par
At liquid helium temperatures, only the lowest crystal field level Z$_1$ is populated, and optical transitions between Z$_1$ and multiple crystal field levels Y$_i$ can be observed. When the electronic Zeeman splitting is much larger than the hyperfine splitting, the transitions between Z$_1$ and Y$_i$ can be split into four transition groups \textit{a,b,c,d} as shown in Fig.~\ref{fig:fig1}(a). The four transition groups are well separated in transition frequency, and each group contains eight hyperfine spin preserving peaks with $\Delta m_I=0$. When the Zeeman splitting and hyperfine splitting become comparable, hyperfine sublevels from different Zeeman branches start to mix.

\begin{figure*}[t]
    \includegraphics[width=\textwidth]{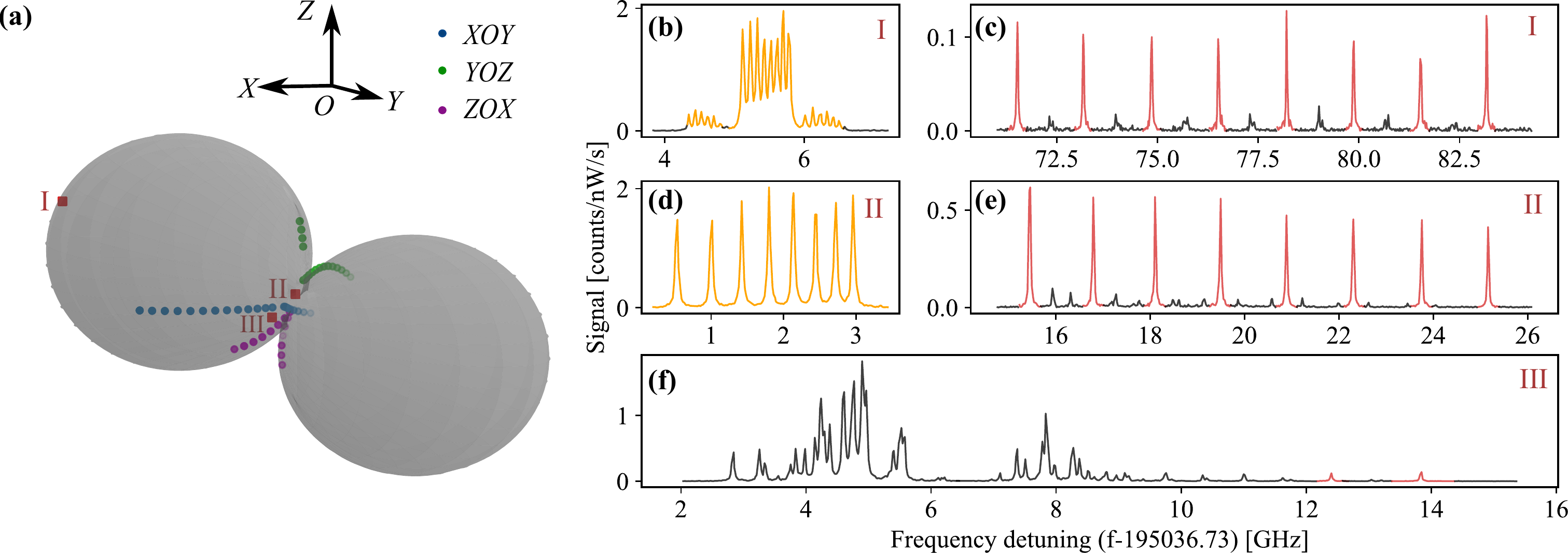}
    \caption{Field rotation Zeeman spectroscopy. (a) A subset of the selected magnetic field directions. The peanut shape contour shows the Zeeman splitting amplitudes of the Z$_1$ level when the magnetic field points in different directions, and the coloured dots on the contour surface denote the selected field directions. Each dot indicates a magnetic field vector direction from the origin $O$ to the dot, and its colour is used to highlight different groups of directions and its brightness comes from the 3D lighting effect. Three sets of spectra are shown as examples in (b)-(f), and the Roman numeral on the top right corner of each figure denotes the magnetic field direction in use and corresponds to the Roman numerals in (a). The coloured spectral peaks in (b)-(f) were used for the spin Hamiltonian fitting.}
    \label{fig:fig2}
\end{figure*}

\section{setup and spectroscopy}
The experimental setup and the device used in this work are illustrated in Fig.~\ref{fig:fig1}(b). The device was a Si fin-field-effect transistor (FinFET) consisting of three terminals and a nanowire channel (\SI{35}{\nano\metre} width $\times$ \SI{80}{\nano\metre} length $\times$ \SI{60}{\nano\metre} height). $^{167}$Er and O were implanted into the device followed by a \SI{700}{\celsius} annealing to repair the implantation damage. The estimated concentration of Er and O in the channel is \SI{1e17}{\per\cubic\centi\meter} and \SI{1e18}{\per\cubic\centi\meter}, respectively. Additionally, the p-type channel had a B doping concentration of \SI{3e18}{\per\cubic\centi\meter}. The device was installed on the cold stage of a liquid helium-free cryostat operating at a base temperature of \SI{3.6}{\kelvin}. The cryostat is equipped with a 6-1-1~\si{\tesla} superconducting vector magnet, which can provide a magnetic field up to \SI{1}{\tesla} in any direction. The device was placed at the centre of the magnet, and the field homogeneity is for the X and Y axes and 0.1\% for the Z axis over a \SI{1}{\cm} diameter of spherical volume.

A fibre-coupled frequency-tunable laser (Pure Photonics PPLC550) was used to excite the optical transitions. The laser light was split into two beams. One beam was sent to a photodetector (PD) and a wavemeter (WM) for power and wavelength monitoring. The specified repeatability of the wavemeter (Bristol 621A-NIR) is \SI{+-6}{\MHz}. The other beam went through a polarisation controller (PC) and optical fibres and reached the cold stage. The light was focused onto the device surface by a two-lens objective\cite{Hogele_Fiber_08}, and the spot size was approximately \SI{2}{\micro\meter}.

Single Er\tplus ion spectra were measured with an optical-electrical hybrid method\cite{Yin_13,hu_er_2022}. The FinFET was biased under a sub-threshold gate voltage\cite{Sellier_Subthreshold_07} so that single quantum dots (QDs) form in the device channel. These QDs can work as sensitive charge sensors and detect the loss or gain of a single electron in its vicinity. After the Er\tplus ion was excited into the \ere excited state by resonant light, it would relax back to the \erg ground state via either a radiative process by photon emission or non-radiative processes. 

Our previous studies have demonstrated that a non-radiative relaxation of an Er\tplus ion can cause a nearby trap to ionise\cite{hu_er_2022}, and afterwards, the trap resets by capturing an electron. These processes can be seen from a typical current-time trace (red) in Fig.~\ref{fig:fig1}(c) measured under resonant illumination. The current through the QD switches between two discrete levels due to ionisation and reset events. In contrast, the black current-time trace 
was measured with a non-resonant laser frequency, and the current stays at the background level. The spectra presented in this work were measured under continuous wave laser excitation, and the spectral signal was defined as the number of ionisation events per unit time per unit power in a long current-time trace. In principle, a longer current-time trace gives a more accurate spectral signal at each frequency, and a finer frequency step size leads to more data points for the peak fitting, but both will lead to longer measurement times. Therefore, a trade-off was made between these two parameters, and a frequency step size of \SI{20}{\MHz} and a trace length of \SI{5}{\s} were used in this study.

\section{Field rotation measurement}
In order to determine the anisotropic spin Hamiltonian tensors, we used the vector magnet to apply fields along different directions while holding the field magnitude fixed at \SI{0.4}{\tesla}. For a low symmetry site, high anisotropy is expected for the $\mathbf{g}$-tensors, and resonant frequencies of the four transition groups vary considerably as the field rotates. Therefore, we started with a preliminary measurement to determine a simplified spin Hamiltonian with only the electronic Zeeman term, which can be used to identify a suitable laser frequency scanning range for each field direction. Firstly, the four transition groups were measured by rotating the field in a circle within the three orthogonal planes ($XOY$,$YOZ$,$ZOX$) of the laboratory frame. Secondly, spectral scans were performed in high magnetic fields along the $Z$ axis to identify the ground electron spin level of Z$_1$ and to assign the \textit{a,b,c,d} optical transition groups to the corresponding Z$_1$ and Y$_i$ electron spin levels. Finally, the fitting results gave the Z$_1$ and Y$_i$ $\mathbf{g}$-tensors, and both turned out to be highly anisotropic with similar orientation. 

The preliminary $\mathbf{g}$-tensors were then used to select the directions for the field rotation measurement. The peanut shape contour in Fig.~\ref{fig:fig2}(a) shows the Zeeman splitting amplitudes of the Z$_1$ level when the magnetic field is along different directions, and the coloured dots on the contour surface denote a subset of the selected field directions while the full set can be found in the supplemental Figure 1. Each dot indicates a magnetic field vector direction from the origin $O$ to the dot, and its colour is used to highlight different groups of directions and its brightness comes from the 3D lighting effect. 

To help the understanding of the field direction selection strategy, three sets of spectra are shown as examples in Figs.~\ref{fig:fig2}(b)-(f). Only the spectral peaks that can be properly assigned to energy levels and show good peak contrast were used for the final spin Hamiltonian fitting, such as the orange and red spectral peaks in Figs.~\ref{fig:fig2}(b)-(f). Multi-peak fitting with Lorentzian functions was used to determine the peak positions. The full width at half maximum (FWHM) of the isolated peaks was approximately \SI{32}{\MHz}\cite{yang_spectral_2022}. Since the spectra were measure with a finite step size of \SI{20}{\MHz}, the apparent peak heights in Figs.~\ref{fig:fig2}(b)-(f) show arbitrary fluctuations and the highest points do not necessarily correspond to the fitted peak heights or centre frequencies.

The two ends of the peanut shape contour in Fig.~\ref{fig:fig2}a correspond to the principal axis of the $\mathbf{g}$-tensor with the largest $\mathbf{g}$ value, defined as $g_z$. A typical set of spectra in this region is shown in Figs.~\ref{fig:fig2}(b),(c), and its field direction is denoted by the red square I in Fig.~\ref{fig:fig2}(a). Specifically, Fig.~\ref{fig:fig2}(b) shows the transition group \textit{b} which comprises eight strong hyperfine peaks corresponding to $\Delta m_I$=0 in the central region and two sets of seven weaker hyperfine peaks corresponding to $|\Delta m_I|=1$ in the two side regions. Figure.~\ref{fig:fig2}(c) shows the transition group \textit{a} with a lower overall signal than group \textit{b}. Group \textit{a} also consists of eight stronger $\Delta m_I$=0 peaks, but the weaker $|\Delta m_I|=1$ peaks distribute between the $\Delta m_I$=0 peaks. These features show up similarly when the magnetic field points in most directions except the waist region of the peanut shape contour, where the two transverse principal axes lie, with smaller $\mathbf{g}$ values, defined as $g_y$ and $g_x$. In this ``transverse'' field region, the Zeeman splitting becomes comparable to the hyperfine splitting. Some $|\Delta m_I|=1$ and $\Delta m_I$=0 hyperfine peaks start to overlap, and some $|\Delta m_I|=1$ peaks become stronger due to the mixing of spin states. These phenomena can be seen from the spectra II and III in Fig.~\ref{fig:fig2}(d)-(f), and the field directions are denoted by the red squares II and III in Fig.~\ref{fig:fig2}(a), respectively.

In total, 181 different field directions were selected into two categories. One category covers the entire space as the magnetic field follows a three-dimensional (3D) spiral path. These field directions can be seen in the supplemental Figure 1, and three of them are denoted by the red squares in Fig.~\ref{fig:fig2}(a). The other category focuses on the transverse field region with three field rotational scans within the three orthogonal planes ($XOY$,$YOZ$,$ZOX$) of the laboratory frame, as indicated by the blue, green, and purple dots in Fig.~\ref{fig:fig2}(a). The spectral information in this region is critical for improving the fitting accuracy of the smaller values in the Zeeman and hyperfine tensors. Due to the level mixing and anti-crossing in this region, intensities of transitions become equivalent and hyperfine peaks are unevenly spaced, as can be seen in spectra II, III\cite{Guillot_hyperfine_06}.

\section{Spin Hamiltonian fitting}
Using the method described above, 5788 hyperfine transition peaks were identified from the 181 spectral scans and were used for the spin Hamiltonian fitting.
For a low symmetry Er site, the principal axes of the $\mathbf{g}$- and $\mathbf{A}$-tensors of different crystal field levels may deviate from each other. Therefore, each tensor has six independent variables, i.e., three principal values and three Euler angles. Euler rotations followed a $z-y'-z''$ sequence. A diagonal matrix $\mathbf{M_p}$ defined in the coordinate system of the principal axes $(x,y,z)$ can be transformed to a matrix $\mathbf{M}$ defined in the laboratory frame $(X,Y,Z)$ by the following relation:
\begin{equation}
	\label{eq:eq2}
	\mathbf{M}=\mathbf{R^T}\cdot \mathbf{M_p} \cdot \mathbf{R},
\end{equation}
where $\mathbf{R}=\mathbf{R}_{z''}(\gamma)\cdot \mathbf{R}_{y^{'}}(\beta) \cdot \mathbf{R}_{z}(\alpha)$, and $\mathbf{R}_{i}$ represents a rotation matrix of an angle ($\alpha, \beta$, or $\gamma$) about the axis $i$. Here $y',z''$  are the new axes after the first and second rotations. 
\par
\begin{figure*}[t]
	\includegraphics[width=0.9\textwidth]{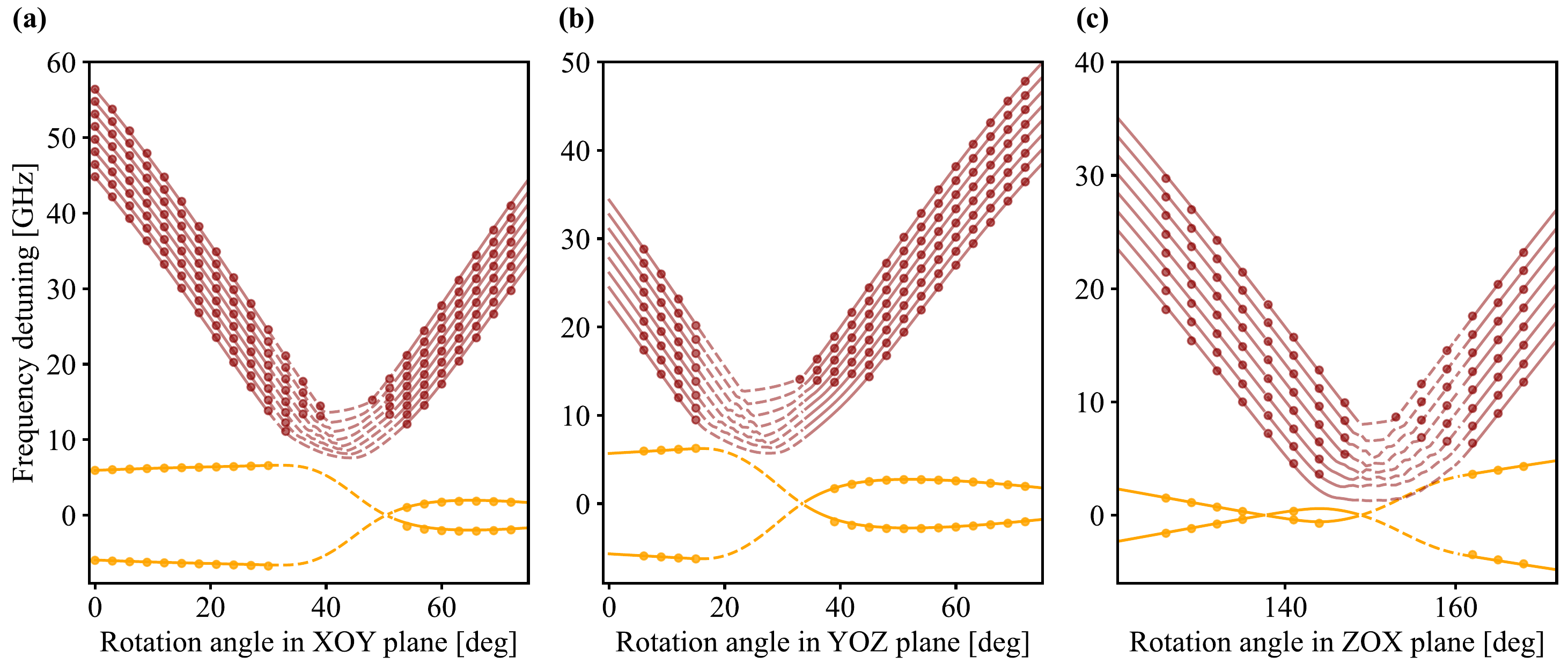} 
	\caption{\label{fig:fig3} Field rotation measurement results and fittings within the three orthogonal planes of the laboratory frame. Measured (coloured dot) and calculated (coloured line) spectral peak positions are plotted as a function of the field rotation angle as the magnetic field rotates within (a) $XOY$ plane, (b) $YOZ$ plane, and (c) $ZOX$ plane. Red dots and lines correspond to the eight $\Delta m_I=0$ transitions in group \textit{a}. Orange dots and lines correspond to groups \textit{b} and \textit{c}. Since the spacing between adjacent hyperfine peaks in groups \textit{b} and \textit{c} is too small to be presented together with group \textit{a}, only the fourth lowest energy transition among the eight $\Delta m_I=0$ hyperfine transitions of group \textit{b} (and \textit{c}) is plotted. Due to state mixing in the anti-crossing regions, the calculated $\Delta m_I=0$ transitions may become weaker than other transitions, e.g., $|\Delta m_I|=1$, so dashed lines are used in these regions.}
\end{figure*}

In principle, the quadrupole interaction term $\mathbf{Q}$ can be determined by the field rotation Zeeman experiment. However, its most significant impact on the identifiable spectral peaks was found to be smaller than \SI{10}{\MHz}. This impact could not be measured accurately due to the \SI{+-6}{\MHz} repeatability of the wavemeter. Therefore, the quadrupole interaction term was not included in the fitting. The final spin Hamiltonian model includes 12 free parameters for each of the Z$_1$ and Y$_i$ levels and another free parameter for the optical transition frequency. The Hamiltonian fitting used a global optimisation technique, basin-hopping, to minimise the root-mean-square deviation ($rmsd$) calculated from the residuals as follows:
\begin{equation}
	\label{eq:eq3}
	rmsd=\sqrt{\frac{1}{N}\sum_{k=1}^{N}(f_{k}^{exp}-f_{k}^{sim})^2},
\end{equation}
where $f_{k}^{sim}$ and $f_{k}^{exp}$ are the simulated and measured centre frequency of spectral peak $k$, and $N$ is the total number of peaks in the fitting.
\begin{table}
	\caption{\label{tab:table2}Principal values and Euler angles for all tensors.}
	\begin{ruledtabular}
		\begin{tabular}{clc}
	        &Principal values&Euler angles\\ 
			\hline
            &$g_{z}=14.846\pm0.028$&$\alpha=137.50\pm0.13$\\
			&$g_{y}=2.38\pm0.18$&$\beta=-66.036\pm0.099$\\ 
			&$g_{x}=0.55\pm0.19$&$\gamma=-155.7\pm2.0$\\
			\cline{2-3}
	        $\textrm{Z}_\textrm{1}({^4}\textrm{I}{_{15/2}})$&$A_{z}=1.558\pm0.019$ & $\alpha=138.77\pm4.79$\\
			&$A_{y}=0.56\pm0.21$ & $\beta=-66.30\pm$6.86\\ 
			&$A_{x}=0.30\pm0.16$ & $ \gamma=-65\pm$47\\
			\hline
	    	&$g_{z}=13.100\pm0.023$&$\alpha=129.74\pm$0.11\\
			&$g_{y}=0.59\pm0.17$&$\beta=-71.87\pm$0.11\\
		    &$g_{x}=0.16\pm0.16$ & $\gamma=-161\pm$25\\
			\cline{2-3}
        	$\textrm{Y}_i({^4}\textrm{I}{_{13/2}})$&$A_{z}=1.773\pm$0.023 &$\alpha=127.39\pm$6.57\\
		    &$A_{y}=0.42\pm0.14$ &$\beta=-72.2\pm$7.5\\
	        &$A_{x}=0.07\pm0.16$ &$\gamma=-101\pm$55\\
		\end{tabular}
	\end{ruledtabular}
\end{table}
The fitting gives a $rmsd$ value of \SI{83}{\mega\hertz} and the fitting parameters are listed in Table~\ref{tab:table2}. All four tensors are highly anisotropic, with one principal value ($g_z$) much larger than the other two ($g_y$ and $g_x$), consistent with low site symmetry. The misalignment between the Z$_1$ and Y$_i$ $\mathbf{g}$-tensors indicates that the Er site has monoclinic (C$_1$) symmetry.

The previously reported cubic Er site in Si has an optical transition frequency of \SI{195.01}{\THz}\cite{przybylinska_optically_1996}, which closely matches the C$_1$ Er site in the present work. Also, the trace of Z$_1$ $\mathbf{g}$-tensor is \num{17.78+-0.40}, which matches the expected value for a $\Gamma_7$ Z$_1$ state. This match reveals that the cubic crystal field approximation holds for this C$_1$ Er site in Si\cite{carey_electron_1999,carey_state_2009}. Overall, these properties suggest that this C$_1$ Er site is likely to be a distorted cubic site.

\begin{figure*}[t]
    \includegraphics[width=0.6\textwidth]{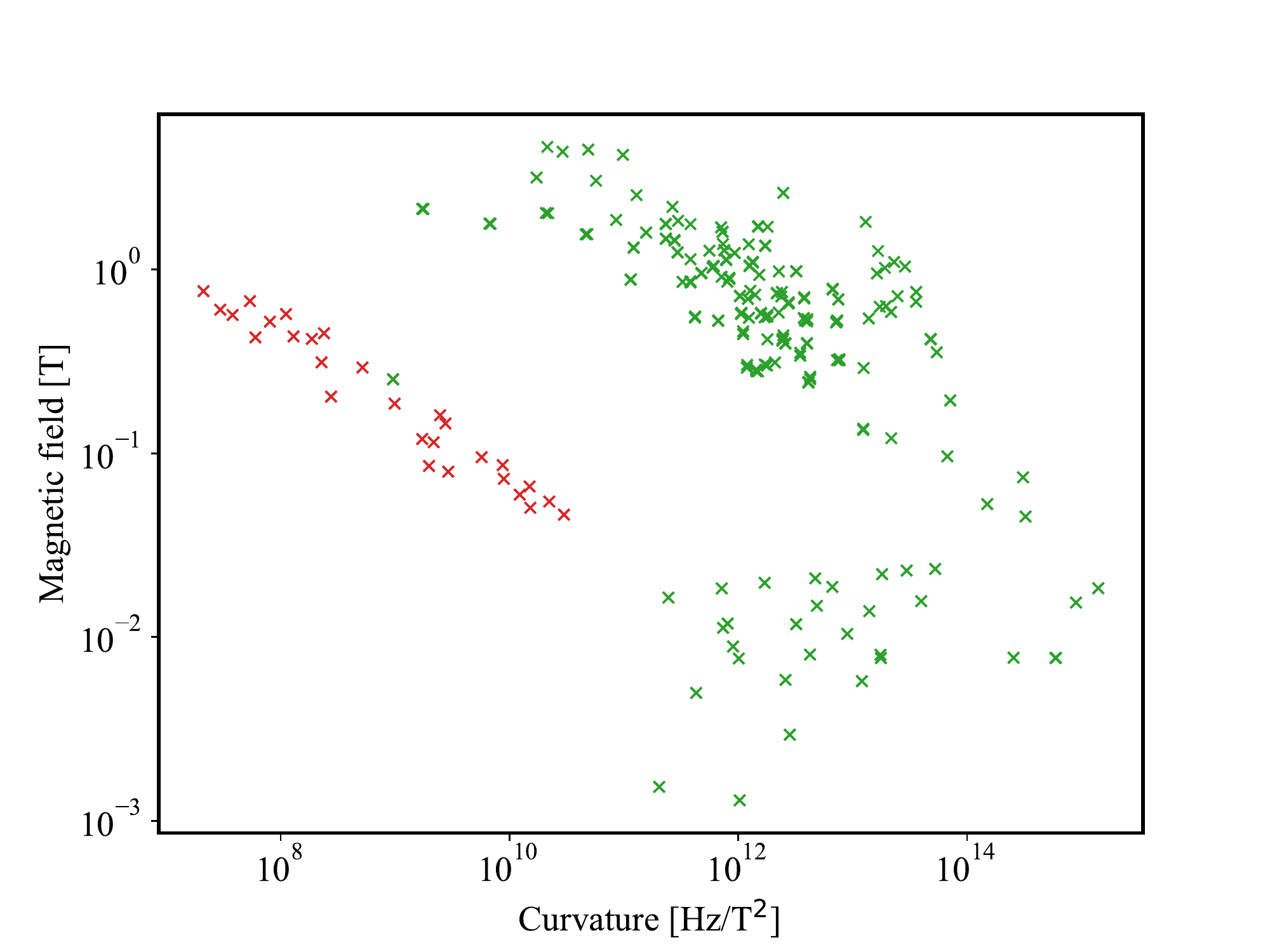}
    \caption{Field strength vs the maximum curvature of the ZEFOZ fields found for the $^{167}$Er\tplus ion. The red and green points correspond to ZEFOZ fields in weakly and strongly mixed regimes, respectively.}
    \label{fig:fig4}
\end{figure*}
Parameter uncertainties in Table~\ref{tab:table2} are calculated using the Markov chain Monte Carlo sampling method. The four tensors' largest principal values ($g_z$, $A_z$) have smaller uncertainties than the transverse principal values ($g_y$, $g_x$, $A_y$, $A_x$) due to the challenges associated with the high anisotropy of the tensors. Firstly, the electron spin quantization axis $\mathbf{\hat{n}=B\cdot g/(|B\cdot g|)}$ remains close to the $z$ direction for a wide range of field directions as a result of the high anisotropy of the $\mathbf{g}$-tensor, except in the transverse field region. This means that the transverse field region is critical for determining the transverse principal values, $g_y$ and $g_x$; however, only a limited number of spectral peaks can be identified in this region due to spin mixing. Secondly, the hyperfine splitting is similarly impacted by the high anisotropy of both $\mathbf{A}$ and $\mathbf{g}$-tensors, as it is primarily determined by $\mathbf{|A\cdot \hat{n}|}$\cite{Chen_hyperfine_18}. Overall, the uncertainties may be underestimated because of the large number of fitting parameters\cite{Chen_hyperfine_18}. In addition, the final position of the device may change slightly during cool-down due to thermal expansion and contraction, and consequently, the actual magnetic field at the Er\tplus ion might deviate from the set field especially for $X$ and $Y$ axes because of the field inhomogeneity.

To evaluate the fitting in the transverse field region, measured and calculated spectral peak positions of groups \textit{a,b,c} from the three in-plane field rotation scans are presented in Fig.~\ref{fig:fig3}. The eight red spectral lines in the upper sections of Figs.~\ref{fig:fig3}(a)-(c) correspond to the eight calculated $\Delta m_I=0$ hyperfine transitions of group \textit{a}, and the red dots represent the measured peaks that can be identified and were used for fitting. The spacing between adjacent hyperfine peaks in groups \textit{b} and \textit{c} is much smaller than group \textit{a}. For clarity, then, only the fourth lowest energy transition among the eight $\Delta m_I=0$ hyperfine transitions of group \textit{b} (and \textit{c}) is plotted in Fig.~\ref{fig:fig3}, and the orange dots and lines represent the measured and calculated peak positions, respectively.

In the region where electronic and hyperfine spins are highly mixed, spectral peaks are generated between sorted hyperfine levels in the Z$_1$ and Y$_i$ states similar to hyperfine transitions outside this region. Due to state mixing, these calculated transitions may become weaker in the anti-crossing region, so dashed lines are used in this region.

The four tensors have very similar $\alpha$ and $\beta$ Euler angles, which determine the direction of the $z$ principal axes of the four tensors. The similarly orientated $\mathbf{A}$- and $\mathbf{g}$-tensors lead to nearly identical hyperfine spectra outside the transverse field region, similar to spectrum I in Fig.~\ref{fig:fig2}(b). This phenomenon can also be seen from the left parts of Figs.~\ref{fig:fig3}(a),(c), even though the three scans in Fig.~\ref{fig:fig3} focus on the transverse field region.

\begin{figure*}[h]
    \includegraphics[width=0.9\textwidth]{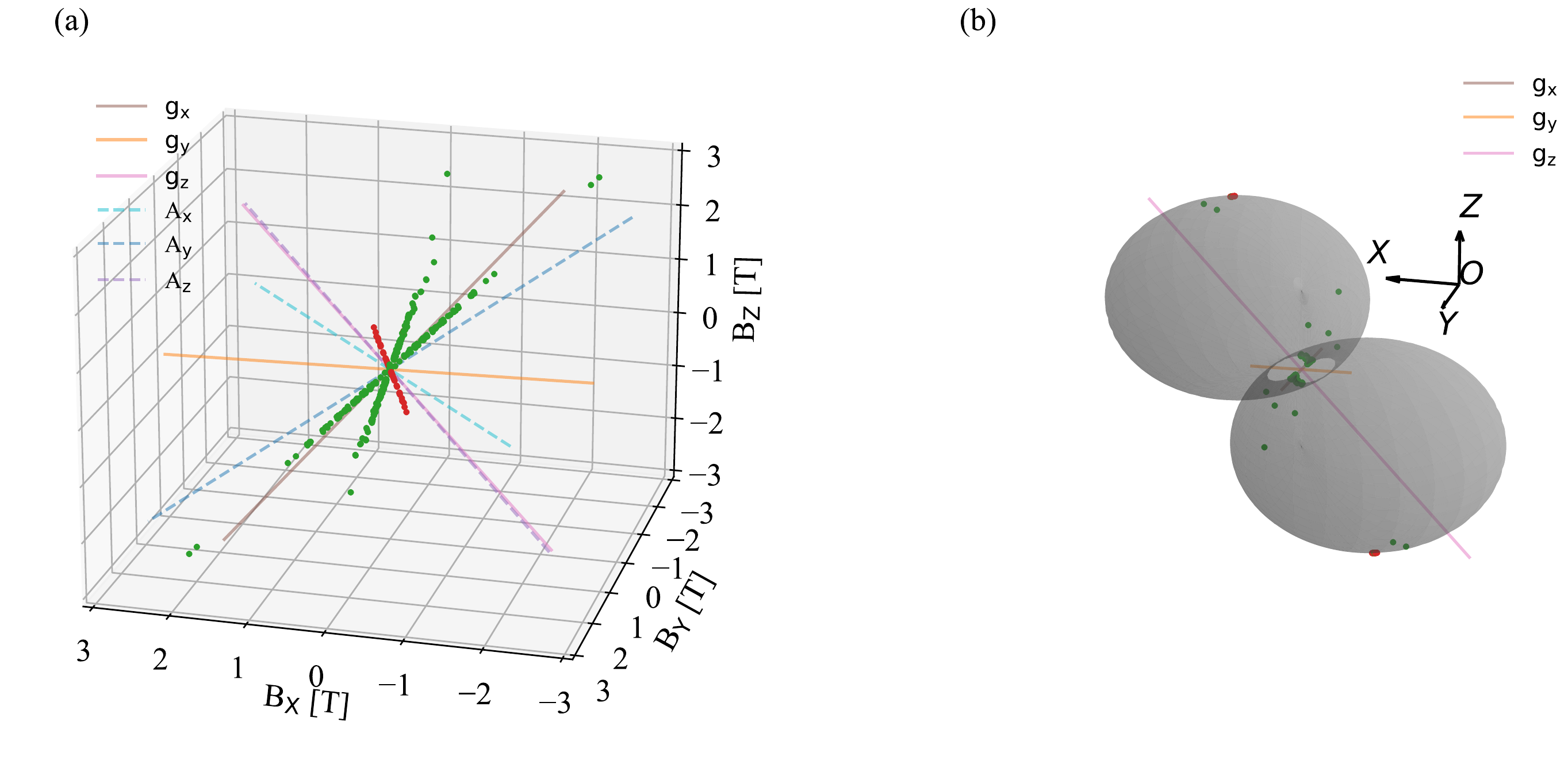}
    \caption{
    3D view of the ZEFOZ fields with a field strength below \SI{5}{\tesla}. (a) ZEFOZ field vectors. Each coloured dot represents a ZEFOZ field vector, and the red and green points represent ZEFOZ fields in weakly and strongly mixed regimes, respectively. The coloured solid and dashed lines indicate the directions of the principal axes of the Z$_1$ level $\mathbf{g}$-tensor and $\mathbf{A}$-tensor, respectively. (b) Directions of the ZEFOZ fields. The peanut shape contour shows the Zeeman splitting amplitudes of the Z$_1$ level when the magnetic field points in different directions. The coloured dots on the contour surface denote the field directions, and each dot indicates a magnetic field vector direction from the origin $O$ to the dot.}
    \label{fig:fig5}
\end{figure*}

In order to locate the ZEFOZ fields for the $^{167}$Er\tplus ion, we follow the calculation procedures described in Ref.\cite{reducing_12} using the fitting results. The hyperfine spin transitions between any any two of the sixteen hyperfine sublevels of Z$_1$ are considered in the calculation. The curvatures at all ZEFOZ fields are calculated and plotted into two groups (red and green) in Fig.~\ref{fig:fig4}. The red group contains field directions close to the largest value principal axis of the Z$_1$ $\mathbf{g}-$tensor. This is a weakly mixed regime, where the electronic Zeeman term dominates, and the electron spin is close to a good quantum number. ZEFOZ fields of the green group are either in the transverse field region of the peanut shape contour or so weak that the electronic Zeeman splitting is smaller than or comparable to the hyperfine splittings. Due to the state mixing, ZEFOZ transitions in this strongly mixed regime tend to have stronger transition strengths than those in the weakly mixed regime. Furthermore, these ZEFOZ fields cluster around particular directions, as shown in Fig.~\ref{fig:fig5}.

The decoherence rate induced by magnetic noise at ZEFOZ fields can be estimated as $S_2(\Delta B)^2$, where $S_2$ is the curvature of ZEFOZ transition and $\Delta B$ is the magnetic field fluctuation. To give an estimation on the decoherence, we consider an experimental condition of sub-Kelvin temperatures and \SI{1}{\tesla} magnetic fields, which can be provided in commercial vector magnet cryogenic systems. The magnetic field fluctuations experienced by $^{167}$Er\tplus ions in natural Si are dominantly due to the flipping of nearby $^{29}$Si nuclear spins and other Er\tplus electron spins. From Monte Carlo simulations\cite{Thesis_Fraval}, $\Delta B$ for an Er doping concentration of \SI{1e17}{\per\cubic\centi\meter} in natural Si is dominated by the flipping of $^{29}$Si nuclear spins and is estimated to be \SI{0.01}{\milli\tesla}. The results suggest that $^{167}$Er\tplus spin coherence times in ZEFOZ fields of about \SI{1}{\tesla} in natural Si can potentially reach \SI{1}{\s} in the strongly mixed regime and \SI{1}{\minute} in the weakly mixed regime. The estimation is indicative, and decoherence suppression is also expected from the frozen core effect\cite{bottger_optical_2006,guillot-noel_direct_2007,Coherence_18}, as the moderate field and low temperature considered here can freeze most electron spins in the crystal. 

\section{Conclusion}
We investigated the Zeeman and hyperfine interactions of the lowest crystal field level Z$_1$ in the \erg manifold and a crystal field level Y$_i$ in the \ere manifold for a single $^{167}$Er\tplus ion in Si. The hyperfine spectra were measured in different magnetic field directions, and the identified peak frequencies were used to fit both the Z$_1$ and Y$_i$ spin Hamiltonians. All four tensors in the spin Hamiltonians are highly anisotropic, with the largest value principal axes in nearly the same direction. The results suggest that the Er\tplus ion occupies a distorted cubic site with monoclinic (C$_1$) symmetry. The ZEFOZ fields calculations for this site suggests $^{167}$Er\tplus spin coherence times above \SI{1}{\s} could be achieved at ZEFOZ fields of about \SI{1}{\tesla} in natural Si.

The accuracy of the spin Hamiltonian parameters could be improved by introducing an on-chip magnetic field sensor. On the other hand, the high anisotropy of the specific Er site leads to larger uncertainties in the transverse properties of the tensors and, consequently, in the transverse ZEFOZ fields. Smaller fitting uncertainties are expected on cubic or other higher symmetry sites, which can be investigated using the field rotation Zeeman method presented in this work.
\begin{acknowledgments}
	The numerical calculations for locating the ZEFOZ fields were performed on the supercomputing system in the Supercomputing Center of University of Science and Technology of China. This work was supported by the National Key R\&D Program of China (Grant No. 2018YFA0306600), Anhui Initiative in Quantum Information Technologies (Grant No. AHY050000), and Anhui Provincial Natural Science Foundation (Grant No. 2108085MA15). We acknowledge the AFAiiR node of the NCRIS Heavy Ion Capability for access to ion-implantation facilities.
\end{acknowledgments}
\section*{Data Availability}
The data that support the findings of this study are available
from the corresponding author upon reasonable request.

\bibliography{erbium_hyperfine}
\end{document}